\documentclass{tlp}
\usepackage{epsfig}
\usepackage[utf8]{inputenc}
\usepackage{multicol}
\usepackage{multirow}
\usepackage{wrapfig}
\usepackage{url}
\newcommand{\verbatimproperties}{\renewcommand{\baselinestretch}{0.85} \small}

\begin{document}

\label{firstpage}

\title[Towards Multi-Threaded Local Tabling Using a Common Table Space]
      {Towards Multi-Threaded Local Tabling \\Using a Common Table Space}

\author[Miguel Areias and Ricardo Rocha]
       {MIGUEL AREIAS and RICARDO ROCHA\\
       CRACS \& INESC TEC, Faculty of Sciences, University of Porto\\
       Rua do Campo Alegre, 1021/1055, 4169-007 Porto, Portugal\\
       \email{\{miguel-areias,ricroc\}@dcc.fc.up.pt}}

\maketitle


\begin{abstract}
  Multi-threading is currently supported by several well-known Prolog
  systems providing a highly portable solution for applications that
  can benefit from concurrency. When multi-threading is combined with
  tabling, we can exploit the power of higher procedural control and
  declarative semantics. However, despite the availability of both
  threads and tabling in some Prolog systems, the implementation of
  these two features implies complex ties to each other and to the
  underlying engine. Until now, XSB was the only Prolog system
  combining multi-threading with tabling. In XSB, tables may be either
  private or shared between threads. While thread-private tables are
  easier to implement, shared tables have all the associated issues of
  locking, synchronization and potential deadlocks. In this paper, we
  propose an alternative view to XSB's approach. In our proposal, each
  thread views its tables as private but, at the engine level, we use
  a \emph{common table space} where tables are shared among all
  threads. We present three designs for our common table space
  approach: \emph{No-Sharing (NS)} (similar to XSB's private tables),
  \emph{Subgoal-Sharing (SS)} and \emph{Full-Sharing (FS)}. The
  primary goal of this work was to reduce the memory usage for the
  table space but, our experimental results, using the YapTab tabling
  system with a local evaluation strategy, show that we can also
  achieve significant reductions on running time. To appear in Theory
  and Practice of Logic Programming.\\
\end{abstract}



\section{Introduction}

Tabling~\cite{Chen-96} is a recognized and powerful implementation
technique that overcomes some limitations of traditional Prolog
systems in dealing with recursion and redundant sub-computations. In a
nutshell, tabling is a refinement of SLD resolution that stems from
one simple idea: save intermediate answers from past computations so
that they can be reused when a \emph{similar call} appears during the
resolution process\footnote{We can distinguish two main approaches to
  determine similarity between tabled subgoal calls:
  \emph{variant-based tabling} and \emph{subsumption-based
    tabling}.}. Tabling based models are able to reduce the search
space, avoid looping, and always terminate for programs with the
\emph{bounded term-size property}~\cite{Chen-96}. Work on tabling, as
initially implemented in the XSB system~\cite{Sagonas-98}, proved its
viability for application areas such as natural language processing,
knowledge based systems, model checking, program analysis, among
others. Currently, tabling is widely available in systems like XSB,
Yap, B-Prolog, ALS-Prolog, Mercury and Ciao.

Nowadays, the increasing availability of computing systems with
multiple cores sharing the main memory is already a standardized,
high-performance and viable alternative to the traditional (and often
expensive) shared memory architectures. The number of cores per
processor is expected to continue to increase, further expanding the
potential for taking advantage of multi-threading support. The ISO
Prolog multi-threading standardization proposal~\cite{Moura-08b} is
currently implemented in several Prolog systems including XSB, Yap,
Ciao and SWI-Prolog, providing a highly portable solution given the
number of operating systems supported by these
systems. Multi-threading in Prolog is the ability to concurrently
perform multiple computations, in which each computation runs
independently but shares the database (clauses).

When multi-threading is combined with tabling, we have the best of
both worlds, since we can exploit the combination of higher procedural
control with higher declarative semantics. In a multi-threaded tabling
system, tables may be either private or shared between threads. While
thread-private tables are easier to implement, shared tables have all
the associated issues of locking, synchronization and potential
deadlocks. Here, the problem is even more complex because we need to
ensure the correctness and completeness of the answers found and
stored in the shared tables. Thus, despite the availability of both
threads and tabling in Prolog compilers such as XSB, Yap, and Ciao,
the implementation of these two features such that they work together
seamlessly implies complex ties to one another and to the underlying
engine. Until now, XSB was the only system combining tabling with
multi-threading, for both private and shared
tables~\cite{Marques-08,Swift-12}. For shared tables, XSB uses a
semi-naive approach that, when a set of subgoals computed by different
threads is mutually dependent, then a \emph{usurpation
  operation}~\cite{Marques-PhD,Marques-10} synchronizes threads and a
single thread assumes the computation of all subgoals, turning the
remaining threads into consumer threads.

The basis for our work is also on multi-threaded tabling using private
tables, but we propose an alternative view to XSB's approach. In our
proposal, each thread has its own tables, i.e., from the thread point
of view the tables are private, but at the engine level we use a
\emph{common table space}, i.e., from the implementation point of view
the tables are shared among all threads. We present three designs for
our common table space approach: \emph{No-Sharing (NS)} (similar to
XSB with private tables), \emph{Subgoal-Sharing (SS)} and
\emph{Full-Sharing (FS)}. Experimental results, using the YapTab
tabling system~\cite{Rocha-05a} with a local evaluation
strategy~\cite{Freire-95}, show that the FS design can achieve
significant reductions on memory usage and on execution time, compared
to the NS and SS designs, for a set of worst case scenarios where all
threads start with the same query goal.

The remainder of the paper is organized as follows. First, we describe
YapTab's table space organization and XSB's approach for
multi-threaded tabling. Next, we introduce our three designs and
discuss important implementation details. We then present some
experimental results and outline some conclusions.


\section{Basic Concepts}

In this section, we introduce some background needed for the following
sections. We begin by describing the actual YapTab's table space
organization, and then we briefly present XSB's approach for
supporting multi-threaded tabling.


\subsection{YapTab's Table Space Organization}

The basic idea behind tabling is straightforward: programs are
evaluated by storing answers for tabled subgoals in an appropriate
data space, called the \emph{table space}. Similar calls to tabled
subgoals are not re-evaluated against the program clauses, instead
they are resolved by consuming the answers already stored in their
table entries. During this process, as further new answers are found,
they are stored in their tables and later returned to all similar
calls.

\begin{figure}[ht]
\centering
\includegraphics[width=8.5cm]{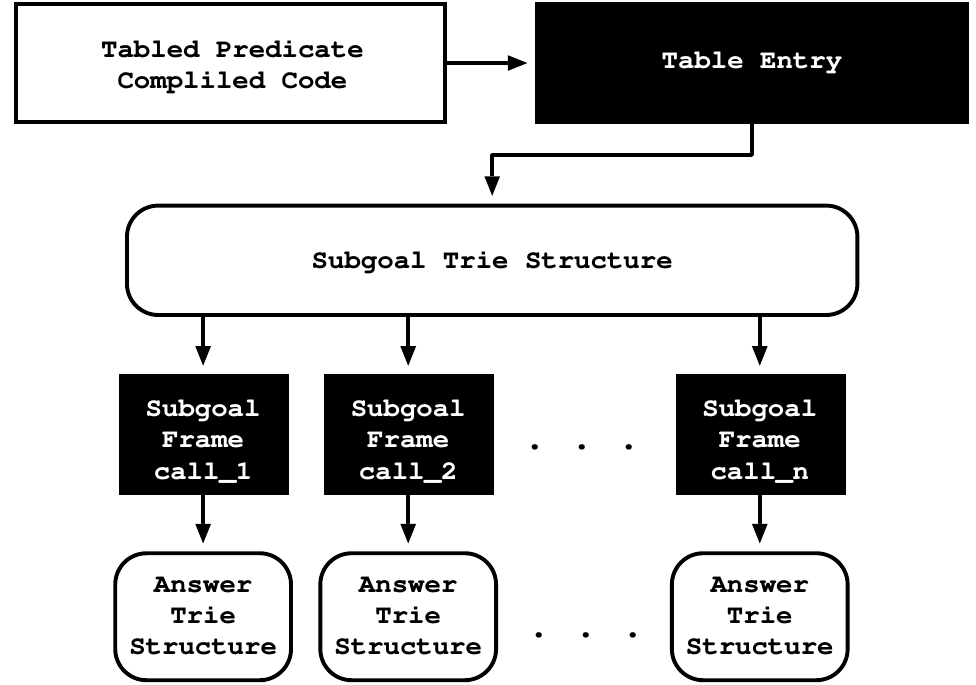}
\caption{YapTab's table space organization}
\label{fig_table_space}
\end{figure}

A critical component in the implementation of an efficient tabling
system is thus the design of the data structures and algorithms to
access and manipulate tabled data. The most successful data structure
for tabling is \emph{tries}~\cite{RamakrishnanIV-99}. Tries are trees
in which common prefixes are represented only once. The trie data
structure provides complete discrimination for terms and permits look
up and possibly insertion to be performed in a single pass through a
term, hence resulting in a very efficient and compact data structure
for term representation. Figure~\ref{fig_table_space} shows the
general table space organization for a tabled predicate in YapTab.

At the entry point we have the \emph{table entry} data structure. This
structure is allocated when a tabled predicate is being compiled, so
that a pointer to the table entry can be included in its compiled
code. This guarantees that further calls to the predicate will access
the table space starting from the same point. Below the table entry,
we have the \emph{subgoal trie structure}. Each different tabled
subgoal call to the predicate at hand corresponds to a unique path
through the subgoal trie structure, always starting from the table
entry, passing by several subgoal trie data units, the \emph{subgoal
  trie nodes}, and reaching a leaf data structure, the \emph{subgoal
  frame}. The subgoal frame stores additional information about the
subgoal and acts like an entry point to the \emph{answer trie
  structure}. Each unique path through the answer trie data units, the
\emph{answer trie nodes}, corresponds to a different answer to the
entry subgoal.


\subsection{XSB's Approach to Multi-Threaded Tabling}
\label{sec_xsb}

XSB offers two types of models for supporting multi-threaded tabling:
\emph{private tables} and \emph{shared tables}~\cite{Swift-12}. 

For private tables, each thread keeps its own copy of the table
space. On one hand, this avoids concurrency over the tables but, on
the other hand, the same table can be computed by several threads,
thus increasing the memory usage necessary to represent the table
space.

For shared tables, the running threads store only once the same table,
even if multiple threads use it. This model can be viewed as a
variation of the \emph{table-parallelism} proposal~\cite{Freire-95},
where a tabled computation can be decomposed into a set of smaller
sub-computations, each being performed by a different thread. Each
tabled subgoal is computed independently by the first thread calling
it, the \emph{generator thread}, and each generator is the sole
responsible for fully exploiting and obtaining the complete set of
answers for the subgoal. Similar calls by other threads are resolved
by consuming the answers stored by the generator thread.

In a tabled evaluation, there are several points where we may have to
choose between continuing forward execution, backtracking, consuming
answers from the table, or completing subgoals. The decision on which
operation to perform is determined by the evaluation strategy. The two
most successful strategies are \emph{batched evaluation} and
\emph{local evaluation}~\cite{Freire-96}. Batched evaluation favors
forward execution first, backtracking next, and consuming answers or
completion last. It thus tries to delay the need to move around the
search tree by batching the return of answers. When new answers are
found for a particular tabled subgoal, they are added to the table
space and the evaluation continues. On the other hand, local
evaluation tries to complete subgoals as soon as possible. When new
answers are found, they are added to the table space and the
evaluation fails. Answers are only returned when all program clauses
for the subgoal at hand were resolved.

Based on these two strategies, XSB supports two types of concurrent
evaluations: \emph{concurrent local evaluation} and \emph{concurrent
  batched evaluation}. In the concurrent local evaluation, similar
calls by other threads are resolved by consuming the answers stored by
the generator thread, but a consumer thread suspends execution until
the table is completed. In the concurrent batched evaluation, new
answers are consumed as they are found, leading to more complex
dependencies between threads. In both evaluation strategies, when a
set of subgoals computed by different threads is mutually dependent,
then a \emph{usurpation operation}~\cite{Marques-10} synchronizes
threads and a single thread assumes the computation of all subgoals,
turning the remaining threads into consumer threads.


\section{Our Approach}
\label{sec_our_approach}

Yap implements a SWI-Prolog compatible multi-threading
library~\cite{Wielemaker-03}. Like in SWI-Prolog, Yap's threads have
their own execution stacks and only share the code area where
predicates, records, flags and other global non-backtrackable data are
stored. Our approach for multi-threaded tabling is still based on this
idea in which each computational thread runs independently. This means
that each tabled evaluation depends only on the computations being
performed by the thread itself, i.e., there isn't the notion of being
a consumer thread since, from each thread point of view, a thread is
always the generator for all of its subgoal calls. We next introduce
the three alternative designs for our approach: \emph{No-Sharing
  (NS)}, \emph{Subgoal-Sharing (SS)} and \emph{Full-Sharing (FS)}. In
what follows, we assume a local evaluation strategy.


\subsection{No-Sharing}

\begin{figure}[ht]
\centering
\includegraphics[width=8.5cm]{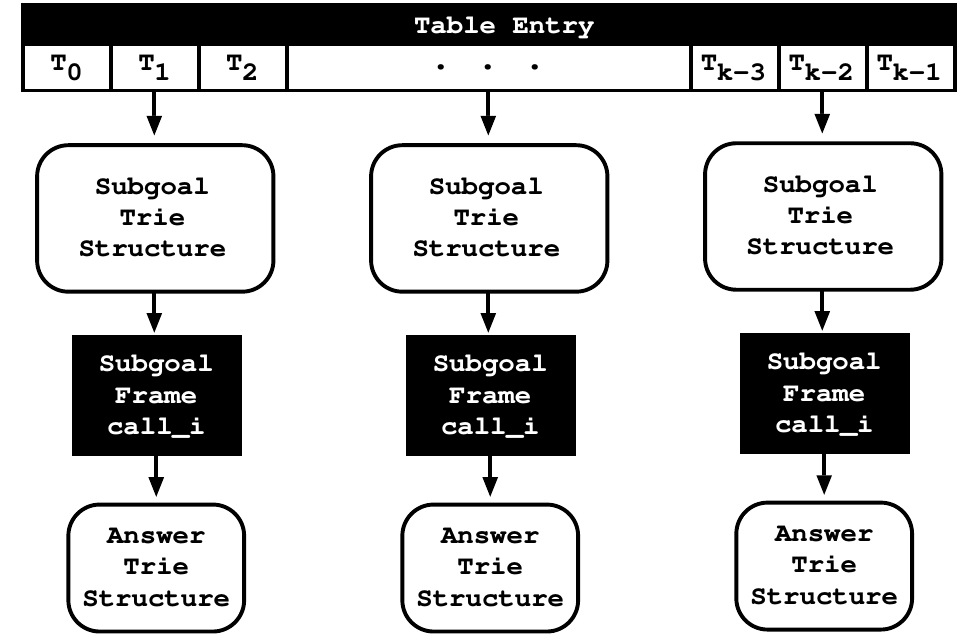}
\caption{Table space organization for the NS design}
\label{fig_no_sharing}
\end{figure}

The starting point of our work is the situation where each thread
allocates fully private tables for each new subgoal called during its
computation. Figure~\ref{fig_no_sharing} shows the configuration of
the table space if several different threads call the same tabled
subgoal $call\_i$. One can observe that the table entry data structure
still stores the common information for the predicate (such as the
arity or the evaluation strategy), and then each thread $t$ has its
own cell $T_t$ inside a \emph{bucket array} which points to the
private data structures. The subgoal trie structure, the subgoal
frames and the answer trie structures are private to each thread and
they are removed when the thread finishes execution.

The memory usage for this design for a particular tabled predicate
$P$, assuming that all running threads $NT$ have completely evaluated
the same number $NS$ of subgoals, is $sizeof(TE_P) + sizeof(BA_P) +
[sizeof(STS_P)+ [sizeof(SF_P) + sizeof(ATS_P)] * NS ] * NT$, where
$TE_P$ and $BA_P$ represent the common table entry and bucket array
data structures, $STS_P$ and $ATS_P$ represent the nodes inside the
subgoal and answer trie structures, and $SF_P$ represents the subgoal
frames.


\subsection{Subgoal-Sharing}

\begin{figure}[ht]
\centering
\includegraphics[width=8.5cm]{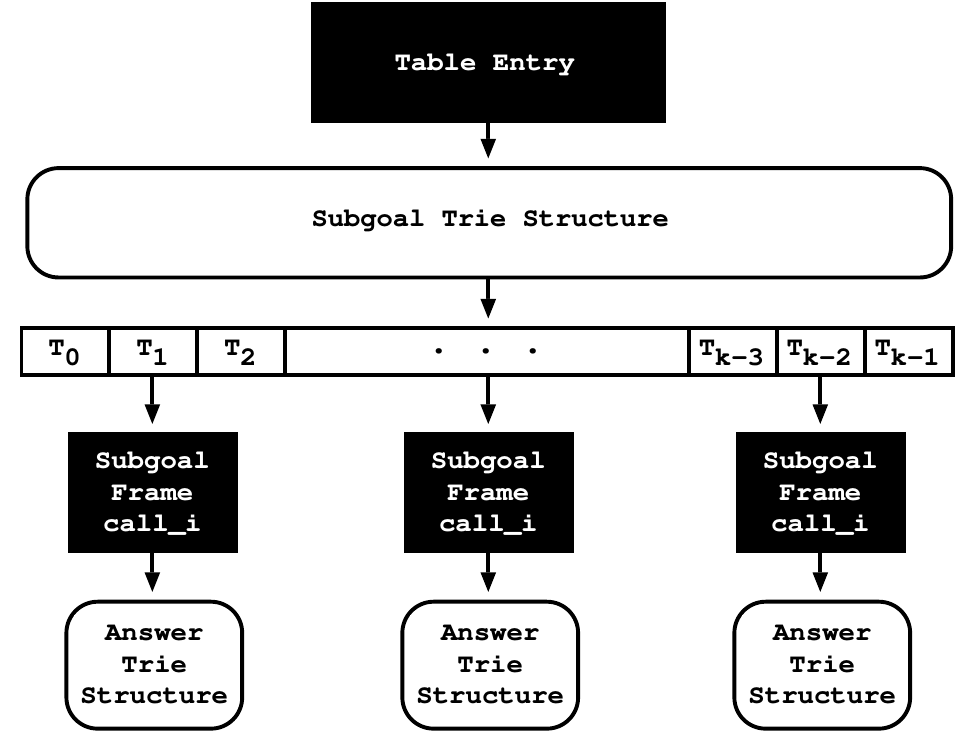}
\caption{Table space organization for the SS design}
\label{fig_subgoal_sharing}
\end{figure}

In our second design, the threads share part of the table
space. Figure~\ref{fig_subgoal_sharing} shows again the configuration
of the table space if several different threads call the same tabled
subgoal $call\_i$. In this design, the subgoal trie structure is now
shared among the threads and the leaf data structures in each subgoal
trie path, instead of pointing to a subgoal frame, they now point to a
bucket array. Each thread $t$ has its own cell $T_t$ inside the bucket
array which then points to a private subgoal frame and answer trie
structure.

In this design, concurrency among threads is restricted to the
allocation of new entries on the subgoal trie structure. Whenever a
thread finishes execution, its private structures are removed, but the
shared part remains present as it can be in use or be further used by
other threads. Assuming again that all running threads $NT$ have
completely evaluated the same number $NS$ of subgoals, the memory
usage for this design for a particular tabled predicate $P$ is
$sizeof(TE_P) + sizeof(STS_P) + [sizeof(BA_P) + [sizeof(SF_P) +
    sizeof(ATS_P)] * NT ] * NS$, where $BA_P$ represents the bucket
array pointing to the private data structures.

\subsection{Full-Sharing}

\begin{figure}[ht]
\centering
\includegraphics[width=10.5cm]{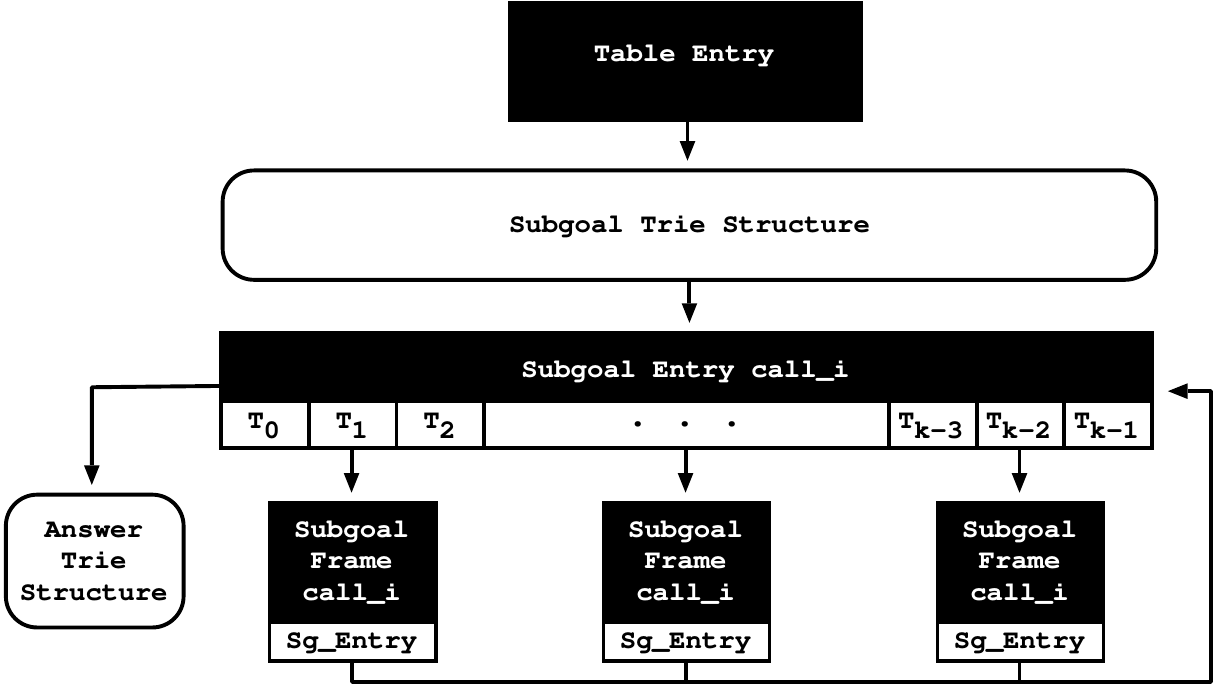}
\caption{Table space organization for the FS design}
\label{fig_full_sharing}
\end{figure}

Our third design is the most sophisticated among
three. Figure~\ref{fig_full_sharing} shows its table space
organization if considering several different threads calling the same
tabled subgoal $call\_i$. In this design, part of the subgoal frame
information (the \emph{subgoal entry} data structure in
Fig.~\ref{fig_full_sharing}) and the answer trie structure are now
also shared among all threads. The previous subgoal frame data
structure was split into two: the \emph{subgoal entry} stores common
information for the subgoal call (such as the pointer to the shared
answer trie structure); the remaining information (the \emph{subgoal
  frame} data structure in Fig.~\ref{fig_full_sharing}) remains
private to each thread.

The subgoal entry also includes a bucket array, in which each cell
$T_t$ points to the private subgoal frame of each thread $t$. The
private subgoal frames include an extra field which is a back pointer
to the common subgoal entry. This is important because, with that, we
can keep unaltered all the tabling data structures that point to
subgoal frames. To access the private information on the subgoal
frames there is no extra cost (we still use a direct pointer), and
only for the common information on the subgoal entry we pay the extra
cost of following an indirect pointer.

Again, assuming that all running threads $NT$ have completely
evaluated the same number $NS$ of subgoals, the memory usage for this
design for a particular tabled predicate $P$ is $sizeof(TE_P) +
sizeof(STS_P) + [sizeof(SE_P) + sizeof(BA_P) + sizeof(ATS_P) +
  sizeof(SF_P)* NT] * NS$, where $SE_P$ and $SF_P$ represent,
respectively, the shared subgoal entry and the private subgoal frame
data structures.

In this design, concurrency among threads now also includes the access
to the subgoal entry data structure and the allocation of new entries
on the answer trie structures. However, this latest design has two
major advantages. First, memory usage is reduced to a minimum. The
only memory overhead, when compared with a single threaded evaluation,
is the bucket array associated with each subgoal entry, and apart from
the split on the subgoal frame data structure, all the remaining
structures remain unchanged. Second, since threads are sharing the
same answer trie structures, answers inserted by a thread for a
particular subgoal call are automatically made available to all other
threads when they call the same subgoal. As we will see in
section~\ref{sec_results}, this can lead to reductions on the
execution time.


\section{Implementation}

In this section, we discuss some low level details regarding the
implementation of the three designs. We begin by describing the
expansion of the table space to efficiently support multiple threads,
next we discuss the locking schemes used to ensure mutual exclusion
over the table space, and then we discuss how the most important
tabling operations were extended for multi-threaded tabling support.


\subsection{Efficient Support for Multiple Threads}

Our proposals already include support for any number of threads
working on the same table. For that, we extended the original table
data structures with bucket arrays. For example, for the NS design, we
introduced a bucket array in the table entry (see
Fig.~\ref{fig_no_sharing}), for the SS design, the bucket array
follows a subgoal trie path (see Fig.~\ref{fig_subgoal_sharing}), and
for the FS design, the bucket array is part of the new subgoal entry
data structure (see Fig.~\ref{fig_full_sharing}).

These bucket arrays contain as much entry cells as the maximum number
of threads that can be created in Yap (currently 1024). However, in
practice, this solution is highly inefficient and memory consuming, as
we must always allocate this huge bucket array even when only one
thread will use it.

\begin{figure}[ht]
\centering
\includegraphics[width=12.5cm]{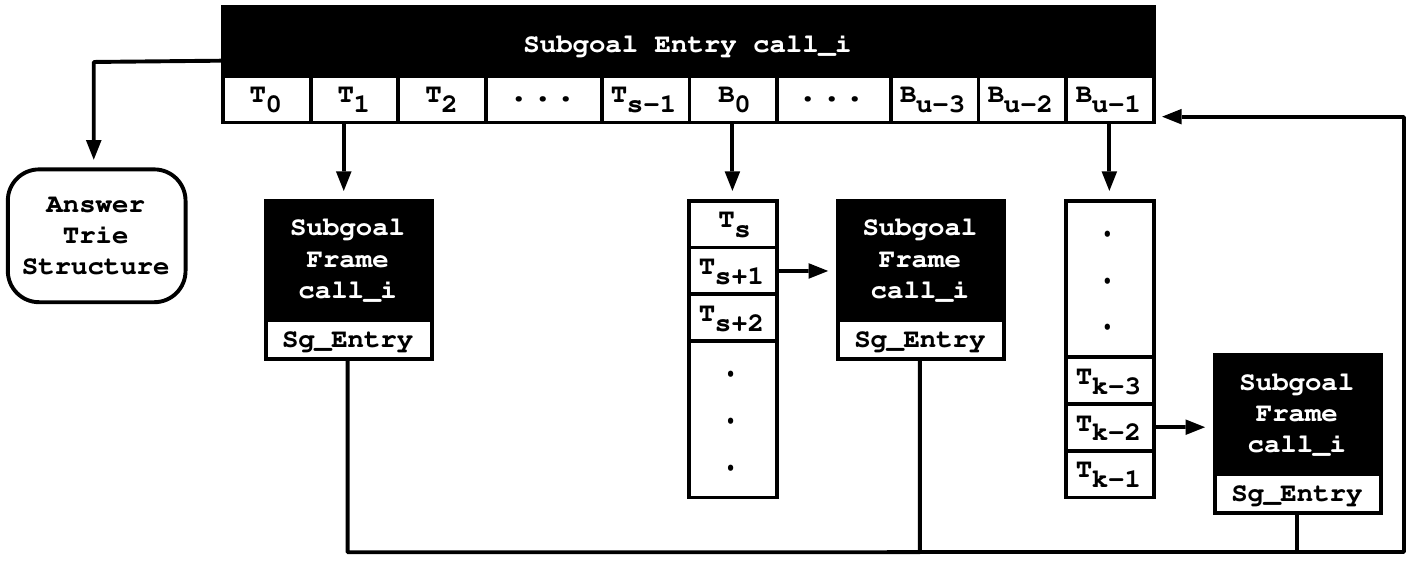}
\caption{Using direct and indirect bucket cells in the FS design}
\label{fig_dynamic_buckets}
\end{figure}

To solve this problem, we introduce a kind of inode pointer structure,
where the bucket array is split into direct bucket cells and indirect
bucket cells. The direct bucket cells are used as before and the
indirect bucket cells are only allocated as needed. This new structure
applies to all bucket arrays in the three
designs. Figure~\ref{fig_dynamic_buckets} shows an example on how this
new structure is used in the FS design.

A bucket array has now two operating modes. If it is being used by a
thread with an identification number $t$ lower than a default starting
size $s$ (32 in our implementation), then the buckets are used as
before, meaning that the entry cell $T_t$ still points to the private
information of the corresponding thread. But now, if a thread with an
identification number equal or higher than $s$ appears, the thread is
mapped into one of the $u$ undirected buckets (entry cells $B_0$ until
$B_{u-1}$ in Fig.~\ref{fig_dynamic_buckets}), which becomes a pointer
to a second level bucket array that will now contain the entry cells
pointing to the private thread information. Given a thread $t$ $(t
\geq s)$, its index in the first and in the second level bucket arrays
is given by the division and the remainder of $(t-s)$ by $u$,
respectively.


\subsection{Table Locking Schemes}

Remember that the SS and FS designs introduce concurrency among
threads when accessing shared resources of the table space. Here, we
discuss how we use locking schemes to ensure mutual exclusion when
manipulating such shared resources.

We can say that there are two critical issues that determine the
efficiency of a locking scheme. One is the \emph{lock duration}, that
is, the amount of time a data structure is locked. The other is the
\emph{lock grain}, that is, the amount of data structures that are
protected through a single lock request. It is the balance between
lock duration and lock grain that compromises the efficiency of
different locking schemes.

The or-parallel tabling engine of Yap~\cite{Rocha-05a} already
implements four alternative locking schemes to deal with concurrent
table accesses: the \emph{Table Lock at Entry Level} (TLEL) scheme,
the \emph{Table Lock at Node Level} (TLNL) scheme, the \emph{Table
  Lock at Write Level} (TLWL) scheme, and the \emph{Table Lock at
  Write Level - Allocate Before Check} (TLWL-ABC) scheme. Currently,
the first three are also available on our multi-threaded
engine. However, in what follows, we will focus our attention only on
the TLWL locking scheme, since its performance showed to be clearly
better than the other two~\cite{Rocha-04a}.

\begin{figure}[ht]
\centering
\includegraphics[width=7.5cm]{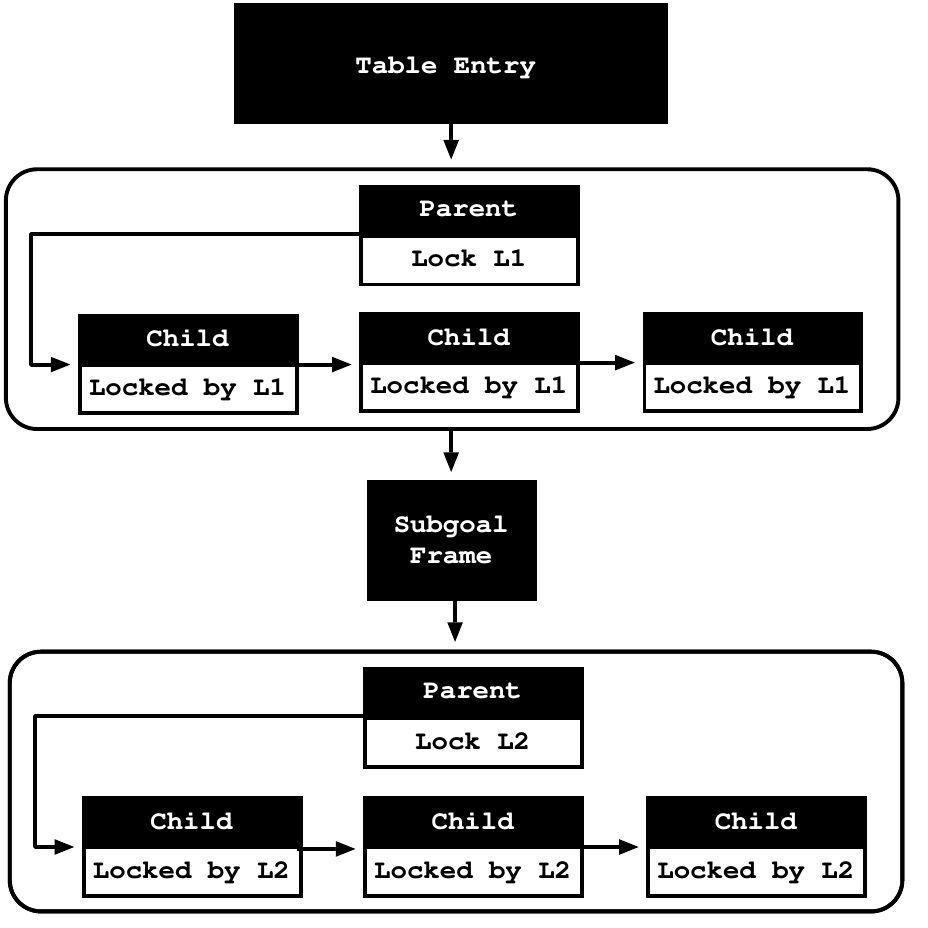}
\caption{The TLWL locking scheme}
\label{fig_locking}
\end{figure}

The TLWL scheme allows a \emph{single writer} per chain of sibling
nodes that represent alternative paths from a common parent node (see
Fig.~\ref{fig_locking}). This means that each node in the
subgoal/answer trie structures is expanded with a \emph{locking field}
that, once activated, synchronizes updates to the chain of sibling
nodes, meaning that only one thread at a time can be inserting a new
child node starting from the same parent node.

With the TLWL scheme, the process of check/insert a term $t$ in a
chain of sibling nodes works as follows. Initially, the working thread
starts by searching for $t$ in the available child nodes (the
non-critical region) and only if the term is not found, it will enter
the critical region in order to insert it on the chain. At that point,
it waits until the lock be available, which can cause a delay
proportional to the number of threads that are accessing the same
critical region at the same time.

In order to reduce the lock duration to a minimum, we have improved
the original TLWL scheme to use \emph{trylocks} instead of traditional
locks. With trylocks, when a thread fails to get access to the lock,
instead of waiting, it returns to the non-critical region, i.e., it
traverses the newly inserted nodes, if any, checking if $t$ was, in
the meantime, inserted in the chain by another thread. If $t$ is not
found, the process repeats until the thread get access to the lock, in
order to insert $t$, or until $t$ be
found. Figure~\ref{fig_trie_check_insert} shows the pseudo-code for
the implementation of this procedure using the TLWL scheme with
trylocks.

\begin{figure}[ht]
{\verbatimproperties
\begin{verbatim}
trie_node_check_insert(term T, parent trie node P)

 1. last_child = NULL  // used to mark the last child to be checked
 2. do {  // non-critical region
 3.   first_child = TrNode_first_child(P)
 4.   child = first_child
 5.   while (child != last_child)  // traverse the chain of sibling nodes ...
 6.     if (TrNode_term(child) == T)  // ... searching for T
 7.       return child
 8.     child = TrNode_sibling(child)
 9.   last_child = first_child
10. } while (! trylock(TrNode_lock(P)))
11. // critical region, lock is set
12. child = TrNode_first_child(P)
13. while (child != last_child)  // traverse the chain of sibling nodes ...
14.   if (TrNode_entry(child) == T)  // ... searching for T
15.     unlock(TrNode_lock(P))  // unlocking before return
16.     return child
17.   child = TrNode_sibling(child)
18. // create a new node to represent T
19. child = new_trie_node(T)
20. TrNode_sibling(child) = TrNode_first_child(P)
21. TrNode_first_child(P) = child
22. unlock(TrNode_lock(P))  // unlocking before return
23. return child
\end{verbatim}}
\caption{Pseudo-code for the trie node check/insert operation}
\label{fig_trie_check_insert}
\end{figure}

Initially, the procedure traverses the chain of sibling nodes, that
represent alternative paths from the given parent node \texttt{P}, and
checks for one representing the given term \texttt{T}. If such a node
is found (line 6) then execution is stopped and the node returned
(line 7). Otherwise, this process repeats (lines 3 to 10) until the
working thread gets access to the lock field of the parent node
\texttt{P}. In each round, the \texttt{last\_child} auxiliary variable
marks the last node to be checked. It is initially set to
\texttt{NULL} (line 1) and then updated, at the end of each round, to
the new first child of the current round (line 9).

Otherwise, the thread gets access to the lock and enters the critical
region (lines 12 to 23). Here, it first checks if $T$ was, in the
meantime, inserted in the chain by another thread (lines 13 to 17). If
this is not the case, then a new trie node representing $T$ is
allocated (line 19) and inserted in the beginning of the chain (lines
20 and 21). The procedure then unlocks the parent node (line 22) and
ends returning the newly allocated child node (line 23).


\subsection{Tabling Operations}

In YapTab, programs using tabling are compiled to include
\emph{tabling operations} that enable the tabling engine to properly
schedule the evaluation process. One of the most important operations
is the \emph{tabled subgoal call}. This operation inspects the table
space looking for a subgoal similar to the current subgoal being
called. If a similar subgoal is found, then the corresponding subgoal
frame is returned. Otherwise, if no such subgoal exists, it inserts a
new path into the subgoal trie structure, representing the current
subgoal, and allocates a new subgoal frame as a leaf of the new
inserted path. Figure~\ref{fig_table_subgoal_call} shows how we have
extended the tabled subgoal call operation for multi-threaded tabling
support.

\begin{figure}[ht]
{\verbatimproperties
\begin{verbatim}
tabled_subgoal_call(table entry TE, subgoal call SC, thread id TI)

 1. root = get_subgoal_trie_root_node(TE, TI)
 2. leaf = check_insert_subgoal_trie(root, SC)
 3. if (NS_design)
 4.   sg_fr = get_subgoal_frame(leaf)
 5.   if (not_exists(sg_fr))
 6.     sg_fr = new_subgoal_frame(leaf)
 7.   return sg_fr
 8. else if (SS_design)
 9.   bucket = get_bucket_array(leaf)
10.   if (not_exists(bucket))
11.     bucket = new_bucket_array(leaf)
12. else if (FS_design)  
13.   sg_entry = get_subgoal_entry(leaf)
14.   if (not_exists(sg_entry))
15.     sg_entry = new_subgoal_entry(leaf)
16.   bucket = get_bucket_array(sg_entry)
17. sg_fr = get_subgoal_frame(bucket)
18. if (not_exists(sg_fr))
19.   sg_fr = new_subgoal_frame(bucket)
20. return sg_fr
\end{verbatim}}
\caption{Pseudo-code for the tabled subgoal call operation}
\label{fig_table_subgoal_call}
\end{figure}

The procedure receives three arguments: the table entry for the
predicate at hand (\texttt{TE}), the current subgoal being called
(\texttt{SC}), and the \emph{id} of the working thread
(\texttt{TI}). The \texttt{NS\_design}, \texttt{SS\_design} and
\texttt{FS\_design} macros define which table design is enabled.

The procedure starts by getting the root trie node for the subgoal
trie structure that matches with the given thread \emph{id} (line
1). Next, it checks/inserts the given \texttt{SC} into the subgoal
trie structure, which will return the leaf node for the path
representing \texttt{SC} (line 2). Then, if the NS design is enable,
it uses the leaf node to obtain the corresponding subgoal frame (line
4). If the subgoal call is new, no subgoal frame still exists and a
new one is created (line 6). Then, the procedure ends by returning the
subgoal frame (line 7). This code sequence corresponds to the usual
tabled subgoal call operation.

Otherwise, for the SS design, it follows the leaf node to obtain the
bucket array (line 9). If the subgoal call is new, no bucket exists
and a new one is created (line 11). On the other hand, for the FS
design, it follows the leaf node to obtain the subgoal entry (line 13)
and, again, if the subgoal call is new, no subgoal entry exists and a
new one is created (line 15). From the subgoal entry, it then obtains
the bucket array (line 16).

Finally, for both SS and FS designs, the bucket array is then used to
obtain the subgoal frame (line 17) and one more time, if the given
subgoal call is new, a new subgoal frame needs to be created (line
19). The procedure ends by returning the subgoal frame (line 20). Note
that, for the sake of simplicity, we omitted some of the low level
details in manipulating the bucket arrays, such as in computing the
bucket cells or in expanding the indirect bucket cels.

Another important tabling operation is the \emph{new answer}. This
operation checks whether a newly found answer is already in the
corresponding answer trie structure and, if not, inserts it. Remember
from section~\ref{sec_xsb} that, with local evaluation, the new answer
operation always fails, regardless of the answer being new or
repeated, and that, with batched evaluation, when new answers are
inserted the evaluation should continue, failing otherwise. With the
FS design, the answer trie structures are shared. Thus, when several
threads are inserting answers in the same trie structure, it may be
not possible to determine when an answer is new or repeated for a
certain thread. This is the reason why the FS design can be only
safely used with local evaluation. We are currently studying how to
bypass this constraint in order to also support the FS design with
batched evaluation.


\section{Experimental Results}
\label{sec_results}

In this section, we present some experimental results obtained for the
three proposed table designs using the TLWL scheme with traditional
locks and with trylocks. The environment for our experiments was a
machine with 4 Six-Core AMD Opteron (tm) Processor 8425 HE (24 cores
in total) with 64 GBytes of main memory and running the Linux kernel
2.6.34.9-69.fc13.x86\_64 with Yap 6.3. To put our results in
perspective, we make a comparison with the multi-threaded
implementation of XSB, version 3.3.6, using thread-private tables.

\begin{figure}[ht]
\centering
\includegraphics[width=10cm]{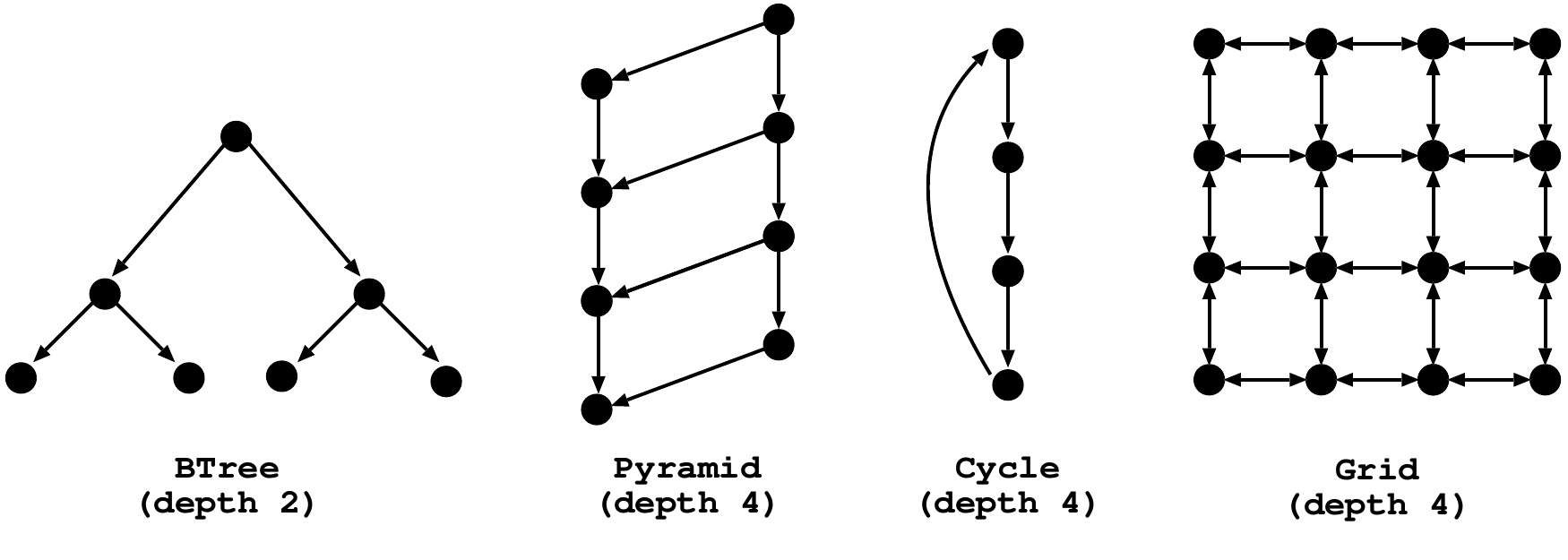}
\caption{Edge configurations}
\label{fig_edge_configurations}
\end{figure}

We used five sets of benchmarks. The \textbf{Large Joins} and
\textbf{WordNet} sets were obtained from the OpenRuleBench
project\footnote{Available from
  \url{http://rulebench.projects.semwebcentral.org}. We also have
  results for the other benchmarks proposed by the OpenRuleBench
  project~\cite{Liang-09} but, due to lack of space, here we only
  include these two sets.}; the \textbf{Model Checking} set includes
three different specifications and transition relation graphs usually
used in model checking applications; the \textbf{Path Left} and
\textbf{Path Right} sets implement two recursive definitions of the
well-known $path/2$ predicate, that computes the transitive closure in
a graph, using several different configurations of $edge/2$ facts
(Fig.~\ref{fig_edge_configurations} shows an example for each
configuration). We experimented the \textbf{BTree} configuration with
depth 18, the \textbf{Pyramid} and \textbf{Cycle} configurations with
depth 2000 and the \textbf{Grid} configuration with depth 35. All
benchmarks find all the solutions for the problem.

Table~\ref{tab_base} shows the execution time, in milliseconds, when
running 1 working thread with local scheduling, for our three table
designs, using the TLWL scheme with traditional locks (columns
\textbf{NS} and \textbf{FS}) and with trylocks (columns
\textbf{SS$_T$} and \textbf{FS$_T$})\footnote{In general, for this set
  of benchmarks, the SS design presented similar results with
  traditional locks and with trylocks and, thus, here we only show the
  results with trylocks.}, and for XSB. In parentheses, it also shows
the respective overhead ratios when compared with the NS design. The
running times are the average of five runs. The ratios marked with
$n.c.$ for XSB mean that we are \emph{not considering} them in the
average results (we opted to do that since they correspond to running
times much higher than the other designs, which may suggest that
something was wrong).

\begin{table}[ht]
\centering
\caption{Execution time, in milliseconds, when running 1 working
  thread with local scheduling, for the NS, SS$_T$, FS and FS$_T$
  designs and for XSB, and the respective overhead ratios when
  compared with the NS design}
\begin{tabular}{lr|rrrr}
\hline\hline
{\bf Bench}
& \multicolumn{1}{c}{\bf NS}
& \multicolumn{1}{c}{\bf SS$_T$}
& \multicolumn{1}{c}{\bf FS}
& \multicolumn{1}{c}{\bf FS$_T$}
& \multicolumn{1}{c}{\bf XSB} \\
\hline
\multicolumn{6}{l}{\bf Large Joins}                     \\
{\bf Join2}    &  3,419 &  3,418 (1.00) &  3,868 (1.13) & 
                           3,842 (1.12) &  3,444 (1.01) \\
{\bf Mondial}  &    730 &    725 (0.99) &    856 (1.17) & 
                             887 (1.21) &  1,637 (2.24) \\
\cline{3-6}\multicolumn{2}{r}{\emph{Average}}
                        &        (1.00) &        (1.15) &
                                 (1.17) &        (1.62) \\
\hline
\multicolumn{6}{l}{\bf WordNet}                         \\
{\bf Clusters} &    789 &    990 (1.26) &    981 (1.24) & 
                             982 (1.24) &    549 (0.70) \\
{\bf Hypo}     &  1,488 &  1,671 (1.12) &  1,728 (1.16) & 
                           1,720 (1.16) & 992,388 ($n.c.$) \\
{\bf Holo}     &    694 &    902 (1.30) &    881 (1.27) &
                             884 (1.27) &    425 (0.61) \\
{\bf Hyper}    &  1,386 &  1,587 (1.15) &  1,565 (1.13) & 
                           1,576 (1.14) &  1,320 (0.95) \\
{\bf Tropo}    &    598 &    784 (1.31) &    763 (1.28) & 
                             762 (1.27) &    271 (0.45) \\
{\bf Mero}     &    678 &    892 (1.32) &    869 (1.28) & 
                             864 (1.27) & 131,830 ($n.c.$) \\
\cline{3-6}\multicolumn{2}{r}{\emph{Average}}
                        &        (1.24) &        (1.23) &
                                 (1.23) &        (0.68) \\
\hline
\multicolumn{6}{l}{\bf Model Checking}                  \\
{\bf IProto}   &  2,517 &  2,449 (0.97) &  2,816 (1.12) &
                           2,828 (1.12) &  3,675 (1.46) \\
{\bf Leader}   &  3,726 &  3,800 (1.02) &  3,830 (1.03) &  
                           3,897 (1.05) & 10,354 (2.78) \\
{\bf Sieve}    & 23,645 & 24,402 (1.03) & 24,479 (1.04) &
                          25,201 (1.07) & 27,136 (1.15) \\
\cline{3-6}\multicolumn{2}{r}{\emph{Average}}
                        &        (1.01) &        (1.06) &
                                 (1.08) &        (1.80) \\
\hline
\multicolumn{6}{l}{\bf Path Left}                       \\
{\bf BTree}    &  2,966 &  2,998 (1.01) &  3,826 (1.29) &  
                           3,864 (1.30) &  2,798 (0.94) \\
{\bf Pyramid}  &  3,085 &  3,159 (1.02) &  3,256 (1.06) &
                           3,256 (1.06) &  2,928 (0.95) \\
{\bf Cycle}    &  3,828 &  3,921 (1.02) &  3,775 (0.99) & 
                           3,798 (0.99) &  3,357 (0.88) \\
{\bf Grid}     &  1,743 &  1,791 (1.03) &  2,280 (1.31) & 
                           2,293 (1.32) &  2,034 (1.17) \\
\cline{3-6}\multicolumn{2}{r}{\emph{Average}}
                        &        (1.02) &        (1.16) &
                                 (1.17) &        (0.98) \\
\hline
\multicolumn{6}{l}{\bf Path Right}                      \\
{\bf BTree}    &  4,568 &  5,048 (1.11) &  5,673 (1.24) & 
                           5,701 (1.25) &  3,551 (0.78) \\
{\bf Pyramid}  &  2,520 &  2,531 (1.00) &  3,664 (1.45) & 
                           3,673 (1.46) &  2,350 (0.93) \\
{\bf Cycle}    &  2,761 &  2,773 (1.00) &  3,994 (1.45) & 
                           3,992 (1.45) &  2,817 (1.02) \\
{\bf Grid}     &  2,109 &  2,110 (1.00) &  3,097 (1.47) & 
                           3,117 (1.48) &  2,462 (1.17) \\
\cline{3-6}\multicolumn{2}{r}{\emph{Average}}
                        &        (1.03) &        (1.40) &
                                 (1.41) &        (0.97) \\
\hline\multicolumn{2}{r}{\emph{Total Average}}
                        &        (1.08) &        (1.21) &
                                 (1.22) &        (1.12) \\
\hline\hline
\end{tabular}
\label{tab_base}
\end{table}

One can observe that, on average, the SS design and the XSB
implementation have a lower overhead ratio (around 10\%) than the FS
and FS$_T$ designs (around 20\%). For the SS and, mainly, for the FS
approaches, this can be explained by the higher complexity of the
implementation and, in particular, by the cost incurred with the extra
code necessary to implement the TLWL locking scheme. Note that, even
with a single working thread, this code has to be executed.

Starting from these base results, Table~\ref{tab_threads} shows the
overhead ratios, when compared with the NS design with 1 thread, for
our table designs and XSB, when running 16 and 24 working threads (the
results are the average of five runs).

\begin{table}[ht]
\centering
\caption{Overhead ratios, when compared with the NS design with 1
  thread, for the NS, SS$_T$, FS and FS$_T$ designs and for XSB, when
  running 16 and 24 working threads with local scheduling (best ratios
  are in bold)} \addtolength{\tabcolsep}{-1.0pt}
\begin{tabular}{lrrrrr|rrrrr}
\hline\hline
\multicolumn{1}{c}{\multirow{2}{*}{\bf Bench}} 
& \multicolumn{5}{c}{\bf 16 Threads}
& \multicolumn{5}{c}{\bf 24 Threads} \\
\cline{2-6}\cline{7-11}
& \multicolumn{1}{c}{\bf NS}
& \multicolumn{1}{c}{\bf SS$_T$}
& \multicolumn{1}{c}{\bf FS}
& \multicolumn{1}{c}{\bf FS$_T$}
& \multicolumn{1}{c}{\bf XSB}
& \multicolumn{1}{c}{\bf NS}
& \multicolumn{1}{c}{\bf SS$_T$}
& \multicolumn{1}{c}{\bf FS}
& \multicolumn{1}{c}{\bf FS$_T$}
& \multicolumn{1}{c}{\bf XSB} \\
\hline
\multicolumn{11}{l}{\bf Large Joins}                                            \\
{\bf Join2}    &      7.96  &      8.05  & \bf{ 3.14} & \bf{ 3.14} &      5.74
               &     24.78  &     24.84  &      3.77  & \bf{ 3.76} &      8.64  \\
{\bf Mondial}  & \bf{ 1.05} &      1.07  &      1.46  &      1.53  &      2.43 
               & \bf{ 1.13} & \bf{ 1.13} &      1.60  &      1.64  &      2.53  \\
\cline{2-11}\emph{Average}
               &      4.51  &      4.56  & \bf{ 2.30} &      2.34  &      4.08
               &     12.96  &     12.98  & \bf{ 2.68} &      2.70  &      5.58  \\
\hline
\multicolumn{11}{l}{\bf WordNet}                                                \\
{\bf Clusters} &      6.29  &      5.61  &      3.92  &      3.94  & \bf{ 2.82} 
               &     12.23  &      8.67  & \bf{ 4.52} &      4.55  &      4.87  \\
{\bf Hypo}     &      5.33  &      5.09  &      4.56  & \bf{ 2.99} &     $n.c.$
               &      9.20  &      8.33  &      5.21  & \bf{ 4.15} &     $n.c.$ \\
{\bf Holo}     &      6.15  &      5.41  &      3.73  &      3.72  & \bf{ 2.77}
               &     10.92  &      9.87  &      4.67  &      4.55  & \bf{ 4.37} \\
{\bf Hyper}    &      8.03  &      7.65  &      3.57  & \bf{ 2.94} &      4.26
               &     21.34  &     16.82  &      4.59  & \bf{ 3.34} &      7.14  \\
{\bf Tropo}    &      6.03  &      4.96  &      3.93  &      3.95  & \bf{ 2.93}
               &     13.46  &      8.44  &      5.64  &      5.68  & \bf{ 4.69} \\
{\bf Mero}     &      4.90  &      4.92  &      3.90  & \bf{ 3.71} &     $n.c.$  
               &      8.93  &      7.96  &      4.59  & \bf{ 4.44} &     $n.c.$ \\
\cline{2-11}\emph{Average}
               &      6.12  &      5.61  &      3.93  &      3.54  & \bf{ 3.19}
               &     12.68  &     10.02  &      4.87  & \bf{ 4.45} &      5.27  \\
\hline
\multicolumn{11}{l}{\bf Model Checking}                                         \\
{\bf IProto}   &      4.15  &      4.20  &      1.60  & \bf{ 1.55} &      1.92 
               &      7.16  &      7.31  &      1.71  & \bf{ 1.63} &      2.14  \\
{\bf Leader}   & \bf{ 1.02} &      1.04  &      1.05  &      1.07  &      2.80
               & \bf{ 1.02} &      1.04  &      1.05  &      1.07  &      2.79  \\
{\bf Sieve}    & \bf{ 1.01} &      1.04  &      1.05  &      1.08  &      1.15 
               & \bf{ 1.02} &      1.04  &      1.06  &      1.08  &      1.15  \\
\cline{2-11}\emph{Average}
               &      2.06  &      2.09  &      1.24  & \bf{ 1.23} &      1.95
               &      3.07  &      3.13  &      1.27  & \bf{ 1.26} &      2.03  \\
\hline
\multicolumn{11}{l}{\bf Path Left}                                              \\
{\bf BTree}    &      9.85  &      9.78  &      6.88  & \bf{ 4.81} &      5.11
               &     25.65  &     25.42  &      8.03  & \bf{ 5.97} &      8.09  \\
{\bf Pyramid}  &      7.67  &      7.79  &      3.74  & \bf{ 3.40} &      4.40 
               &     24.92  &     24.88  &      5.86  & \bf{ 4.48} &      7.02  \\
{\bf Cycle}    &      7.32  &      7.38  &      3.73  & \bf{ 3.25} &      4.36 
               &     22.39  &     23.05  &      5.95  & \bf{ 4.08} &      6.99  \\
{\bf Grid}     &      5.99  &      6.00  &      3.77  &      3.15  & \bf{ 2.41}
               &     19.82  &     19.80  &      4.65  & \bf{ 4.46} &      5.30  \\
\cline{2-11}\emph{Average}
               &      7.71  &      7.74  &      4.53  & \bf{ 3.65} &      4.07
               &     23.20  &     23.29  &      6.12  & \bf{ 4.75} &      6.85  \\
\hline
\multicolumn{11}{l}{\bf Path Right}                                             \\
{\bf BTree}    &     13.82  &     13.13  &     10.57  & \bf{ 5.54} &      6.33
               &     29.53  &     27.36  &     10.16  & \bf{ 6.76} &     10.38  \\
{\bf Pyramid}  &     17.09  &     17.00  &     14.85  &      8.15  & \bf{ 5.94}
               &     46.25  &     45.31  &     10.86  &     10.42  & \bf{10.31} \\
{\bf Cycle}    &     17.96  &     18.17  &     17.05  &      8.36  & \bf{ 6.63}
               &     47.89  &     47.60  &     11.49  & \bf{10.76} &     10.99  \\
{\bf Grid}     &      9.52  &      9.48  &      7.13  &      5.53  & \bf{ 3.75}
               &     26.58  &     27.80  &      7.50  &      6.96  & \bf{ 6.41} \\
\cline{2-11}\emph{Average}
               &     14.60  &     14.44  &     12.40  &      6.90  & \bf{ 5.66}
               &     37.56  &     37.02  &     10.00  & \bf{ 8.73} &      9.52  \\
\hline\emph{Total Average}
               &      7.43  &      7.25  &      5.24  & \bf{ 3.78} &      3.87
               &     18.64  &     17.72  &      5.42  & \bf{ 4.73} &      6.11  \\
\hline\hline
\end{tabular}
\label{tab_threads}
\end{table}

In order to create a worst case scenario that stresses the trie data
structures, we ran all threads starting with the same query goal. By
doing this, it is expected that they will access the table space, to
check/insert for subgoals and answers, at similar times, thus causing
a huge stress on the same critical regions. In particular, for this
set of benchmarks, this will be specially the case for the answer
tries (and thus, for the FS and FS$_T$ designs), since the number of
answers clearly exceeds the number of subgoals. Analyzing the general
picture of Table~\ref{tab_threads}, one can observe that, on average,
the NS and SS$_T$ designs show very poor results for 16 and 24
threads. In particular, these bad results are more clear in the
benchmarks that allocate a higher number of trie nodes. The
explanation for this is the fact that we are using Yap's memory
allocator, that is based on Linux system's \emph{malloc}, which can be
a problem, when making a lot of memory requests, since these requests
require synchronization at the low level implementation.

For the FS and FS$_T$ designs, the results are significantly better
and, in particular for FS$_T$, the results show that its trylock
implementation is quite effective in reducing contention and,
consequently, the running times for most of the experiments.
Regarding XSB, for 16 threads, the results are similar to the FS$_T$
design (3.87 for XSB and 3.78 for FS$_T$, on average) but, for 24
threads, the FS$_T$ is noticeable better (6.11 for XSB and 4.73 for
FS$_T$, on average). These results are more important since XSB shows
base execution times (with 1 thread) lower than FS$_T$ (please revisit
Table~\ref{tab_base}) and since FS$_T$ also pays the cost of using
Yap's memory allocator based on Linux system's \emph{malloc}.

We can say that there are two main reasons for the good results of the
FS design. The first, and most important, is that the FS design can
effectively reduce the memory usage of the table space, almost
linearly in the number of threads\footnote{We have experimental
  results confirming the memory usage formulas introduced on
  section~\ref{sec_our_approach} but, due to lack of space, we are not
  including them here.}, which has the collateral effect of also
reducing the impact of Yap's memory allocator. The second reason is
that, since threads are sharing the same answer trie structures,
answers inserted by a thread are automatically made available to all
other threads when they call the same subgoal. We observed that this
collateral effect can also lead to unexpected reductions on the
execution time.


\section{Conclusions}

We have presented a new approach to multi-threaded tabled evaluation
of logic programs using a local evaluation strategy. In our proposal,
each thread views its tables as private but, at the engine level, the
tables are shared among all threads. The primary goal of our work was,
in fact, to reduce the memory table space but, our experimental
results, showed that we can also significantly reduce the running
times. Since our implementation achieved very encouraging results on
worst case scenario tests, it should keep at least the same level of
efficiency on any other tests. Moreover, we believe that there is
still considerable space for improvements, mainly related to the
low-level issues of Yap's memory allocator for multi-threaded
support. The goal would be to implement strategies that pre-allocate
bunches of memory in order to minimize the performance degradation
that the system suffers, when it is exposed to simultaneous memory
requests made by multiple threads. Further work will also include
extending the FS design to support batched evaluation.


\section*{Acknowledgments}

This work is partially funded by the ERDF (European Regional
Development Fund) through the COMPETE Programme and by FCT (Portuguese
Foundation for Science and Technology) within projects HORUS
(PTDC/EIA-EIA/100897/2008) and LEAP (PTDC/EIA-CCO/112158/2009). Miguel
Areias is funded by the FCT grant SFRH/BD/69673/2010.

\bibliographystyle{acmtrans}
\bibliography{references}


\end{document}